\documentclass[12pt]{article}

\usepackage{graphicx}

\begin{document}

\centerline{\Large \bf Properties of collapse dynamics }
\centerline{\Large \bf in relativistic theory of gravitation }
\centerline{\Large \bf in the case of smooth initial matter distributions }
\vskip 5 mm 

\centerline{\it K.V.Antipin $^{1,2}$, A.I.Dubikovsky $^{1,2}$, P.K.Silaev $^{1,2}$}
\vskip 5 mm 

\centerline{\it $^{1}$ M.V.Lomonosov  Moscow State University,}
\centerline{\it N.N.Bogoliubov Institute for Theoretical Problems of Microphysics,}
\centerline{\it Leninskie Gory,  Moscow, 119991,  Russia}

\centerline{\it $^{2}$ M.V.Lomonosov  Moscow State University, Department of Physics,}
\centerline{\it Leninskie Gory, Moscow, 119991, Russia}
\centerline{\it e-mail: dubikovs@bog.msu.ru}
\vskip 5 mm 

\begin{abstract}
We use both numerical and analytical approaches to study the dynamics of the gravitational collapse in the framework of the relativistic theory of gravitation~(RTG). 
We use various equations of state for the collapsing matter and relatively realistic initial conditions with smooth matter distribution, which corresponds to static solution for the given equation of state. We also obtain results concerning the influence of the graviton mass on
the properties of static solutions. 
We specify several characteristics of the process of the collapse, in particular, we determine the dependence of the turning point time (when contraction is replaced by inflation) on the graviton mass. We also study the influence of non-zero pressure on the dynamics of the collapse.
\end{abstract}


\section{Introduction}
As far back as the 1930s it was shown that the evolution of a dust sphere according to the equations of the general theory of relativity leads to the formation of a black hole~\cite{oppen1, oppen2}. In the framework of the relativistic theory of gravitation~\cite{kni1, kni2} the non-zero graviton mass changes the situation drastically.
Although in Newtonian and post-Newtonian range the mass of the graviton doesn't have much effect on properties of the static solution, in the case of strong fields its impact is highly significant.
A general analysis of this impact on the static spherically symmetric solution has been made in~\cite{stat}. It has been shown that these solutions bear no resemblance to black holes. An analysis of the dynamics of spherically symmetric distribution of cold dust has been performed in~\cite{dyn} where it has been shown that, instead of collapsing into a black hole, a dust sphere starts to pulsate in the course of evolution.

In this paper we will perform numerical analysis of the dynamics of collapse for several specific initial configurations. Firstly, this will allow us to obtain more detailed description of the evolution of the system considered. Secondly, we will be able to verify an intuitively obvious assumption that substitution of the cold dust model by a more realistic one with non-zero pressure will not lead to any qualitative changes. The presence of non-zero pressure  will only slow down~(or even stop) the process of collapse. Thirdly, we will examine the gravitational collapse in the massless limit of the RTG and compare these results with those obtained in the general relativity. And finally, on the basis of obtained numerical results we will derive rigorous analytical results concerning the time of reaching the turning point in massive RTG. 

The paper is structured as follows. In section 1 we investigate static spherically symmetric solutions that further will be used as initial configurations for the process of collapse. We study cases of 
massive and massless RTG and specify the influence of the graviton mass on the properties of the static solution. In section 2 we consider the process of collapse in massless RTG. In section 3 we consider the process of collapse in massive RTG and determine how the turning point time depends on the mass of the graviton.

\section{Initial static configurations}

Lets us consider static configurations that corresponds to a given equation of state for the matter and various values of graviton mass (including the case of massless graviton). We assume that the matter equation of state takes the form:
\begin{equation}
p(r)=C e^\alpha(r). 
\label{mater}
\end{equation}
We perform calculations for various values of constants $C$ and $\alpha$. 
For our purposes it is convenient to parameterize the metric coefficients in the following way:
$$ g_{00}=u(r)=v(r)K^4(r)/L^4(r), \qquad g_{22}= w(r)=K^2(r), $$ 
$$ g_{11}=v(r)=2K(r)K'(r)/r-K^2(r)L'(r)/(rL(r)), \qquad g_{33}= w(r)\sin^2(\theta). $$
Such a parameterization permits us to simplify calculations, because equations $D_i\tilde g^{ij}=0$ will be satisfied automatically. 
The system of equations for $K(r)$, $L(r)$, $p(r)$ takes the form (for short, we omit the argument $r$ in these functions): 
\begin{equation}
K''=\Big[\big(K^4 L' -2 K^3 K' L\big) \big(e- \mu^2-p\big)+2 K^2 L' \big(\mu^2
r^2/2-1\big)+
\label{static}
\end{equation}
$$
+4 K K' \big(-L \mu^2 r^2/2+L+L' r\big)-6
K'^2 L r\Big]/\big(2 K L r\big),
$$
$$L''=\Big[ 2 K^6 K' L L' (e+ \mu^2+3 p)-2 K^7
L'^2 (\mu^2/2+p)+
$$
$$
+K^5
\big(-4 e (K')^2 L^2-4 (K')^2 L^2 p+L L'
\mu^2 r/2+2 (L')^2 (\mu^2 r^2/2-1)\big)-
$$
$$
-2 K^4 K' \big(L^2 \mu^2 r/2+L
L' (2 \mu^2 r^2/2-3)-5 (L')^2 r\big)-
$$
$$-2 K^3
(K')^2 L (2 L+11 L' r)+12 K^2 (K')^3 L^2 r-
$$
$$-K L^5 L'
\mu^2/2 r+2 K' L^6 \mu^2/2 r\Big]/\big(2 K^4 K' L r\big),
$$
$$
p'=\Big[(e+p) \Big( \big(2 K^6 L' -4 K^5
K' L \big)(\mu^2/2+p) - K^4 L \mu^2 r/2
$$
$$
+2 K^3\big(2 K' L-KL'\big) (\mu^2
r^2/2-1)+$$ 
$$+2 K^2 (K')^2 L r+L^5 \mu^2 r/2\Big)\Big]/\big(4 K^3 K' L
r\big).
$$
For the case of massless RTG one should set the graviton mass $\mu$ equal to zero.
We solve this system of equations with boundary conditions $K(0)=0$, $L(0)=0$, $p(\infty)=0$, $K'(\infty)=1$, $L'(\infty)=1$. Of course, in numerical calculations the 
infinity ($r\to\infty$) will be replaced by sufficiently large $r$, and exact boundary conditions
should be replaced by expansions into series in powers of $1/r$ and exponential factors $\exp(-\mu r)$.

\begin{figure}
\includegraphics{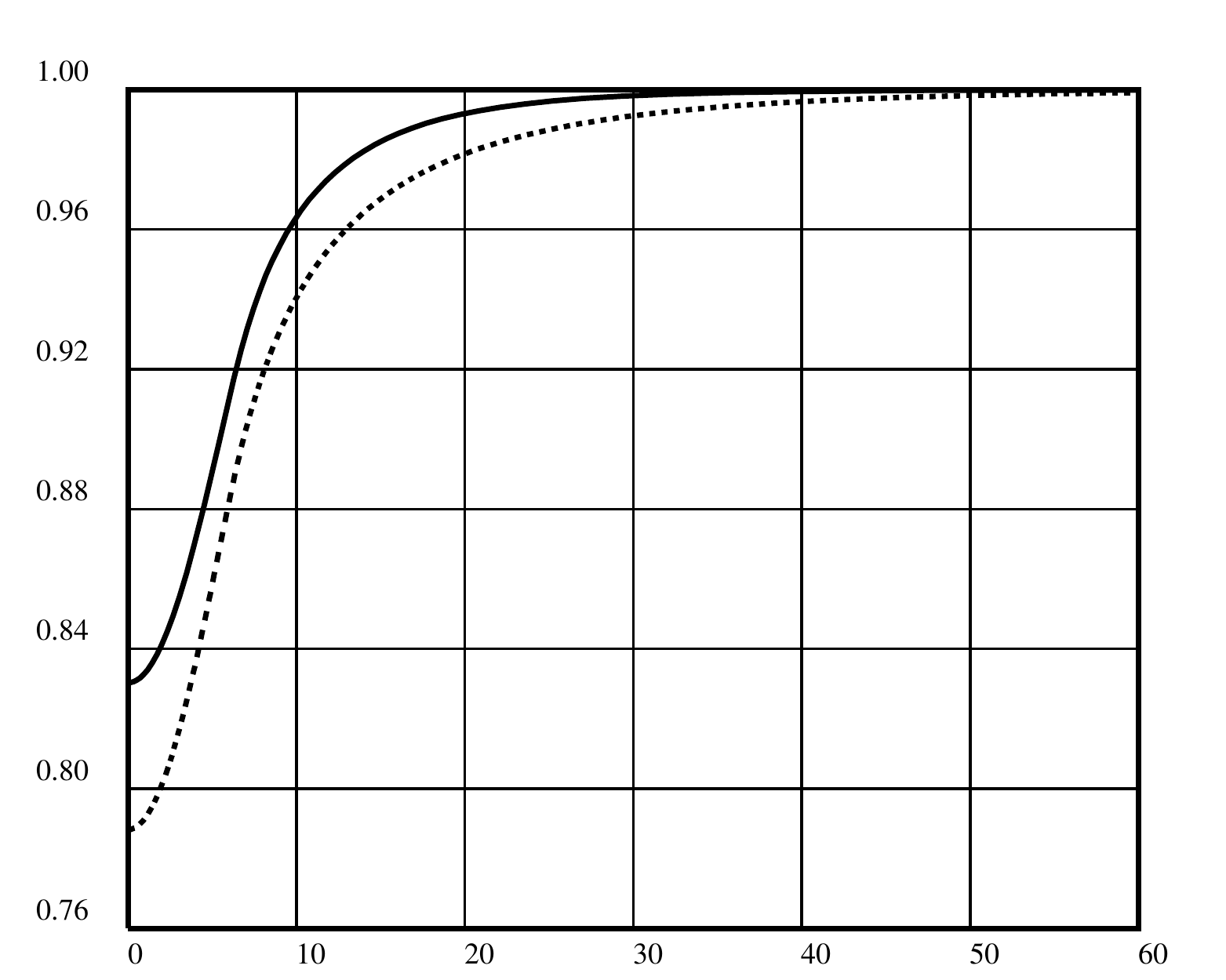}
\caption{Two different static solutions for identical matter state equation and identical solution mass, but different graviton mass: solid line --- metric coefficient $g_{00}(r)$ for the graviton mass $\mu$;
dashed line --- metric coefficient $g'_{00}(r)$ for the graviton mass $\mu'<\mu$. 
}
\end{figure}

For the case of massless RTG it should be noted, that in the region where density is approximately equal to null, the solution doesn't coincide with the simplest static solution of the massless RTG in vacuum
$$ u=( r-m)/( r+m),\qquad v=( r+m)/( r-m),\qquad w=( r+m)^2. $$
The latter, in turn, corresponds to the simplest solution for harmonic radial coordinate 
(see \cite{fock}):
$$r_1(\tilde r)=\tilde r-m,$$ where $\tilde r$ is the standard radial coordinate for 
Schwarzschild 
solution 
$$ g_{00}=1-2m/\tilde r,\qquad g_{11}=-1/g_{00},\qquad g_{22}=-(\tilde r)^2. $$
This fact has already been mentioned in \cite{avak},\cite{genk}. It was proved in \cite{avak},\cite{genk}, that for the case of harmonic coordinates 
internal and external solutions are consistent only for the general solution for harmonic radial coordinate, that includes also the second (logarithmic) solution:
$$ r_2(\tilde r)= 2m+(\tilde r-m)\log(1-2m/\tilde r).$$ 
Our numerical solution shows that at large distances $u(r)$ behaves in such a way that the coefficient of $1/r^3$ is not equal to $-2m^3$, and this means that the coefficient of the second~(logarithmic) solution $ r_2(\tilde r)$ is not equal to 0. 
In fact, it is an obvious result --- the solution at $r\to 0$ is determined by three arbitrary constants~($K(0)=L(0)=0$), so that at $r\to\infty$ we need two arbitrary constants --- the coefficient of $1/r$, that is related to the mass of the solution, and the coefficient of $1/r^3$, that is related to the coefficient of the logarithmic term $ r_2(\tilde r)$.

The main problem for the massive RTG is that we have to perform our numerical simulation at totally unrealistic values of the graviton mass. The known upper limit on the graviton mass is~$m_g<3\cdot 10^{-66}g$ (so that $\mu<8\cdot 10^{-29}\hbox{cm}^{-1}$). Consequently, the mass of the solution should differ from the graviton mass by at least 100 orders of magnitude. This leads to the following technical difficulty: in order to study the influence of the graviton mass on the solution we need to use an accuracy of at least 160 digits, which makes the calculations extremely slow.
For this reason we choose a model in which the graviton mass is only a few orders of magnitude less than the mass of the solution.
The main conclusion concerning the role of the non-zero graviton mass on the static solution for the given equation of state for the matter (besides the obvious exponential behavior of metric coefficient at large distances \cite{kni2}) is that the increasing of graviton mass leads to some increasing of~$u(0)$ --- the larger is $\mu$, the closer to 1 is the value of $u(0)$ (see Fig. 1).
Generally, the initial static configurations for both massive and massless cases appears to be the
smooth matter distributions. Formally in such a distribution there is no ``body boundary'' and ``external'' and ``internal'' solutions, but it always possible to specify the sphere, that includes almost all matter of the solution. In Fig. 1. this sphere approximately corresponds to the region of rapid increasing of metric coefficient $g_{00}(r)$.

\section{Dynamics in Massless Theory}

The initial condition for the collapse dynamic is the static solution, obtained in the previous section. So we assume $u(t=0,r)=u(r)$, $v(t=0,r)=v(r)$, etc. 
We start the evolution of the system by modifying the equation of state~(\ref{mater}) by decreasing the coefficient $C$~(this is in some sense a simulation of nuclear fuel burnup).
For our purposes it is convenient to change~$C$ smoothly:
\begin{equation} C(t)=C_1+(C_0-C_1)\exp(-\beta t^2). \label{zapu}
\end{equation}
Here $\beta$ --- some constant, $C_0$ --- the initial value, that corresponds to initial static configuration, $C_1$ --- the final value of the coefficient (in the case of cold dust $C_1=0$). 

The dynamics of the system is determined by the dynamics of the metric and the dynamics of matter.
Due to our parameterization (see the previous section) the problem of the metric dynamics is reduced to solving the equation for~$K(t,r)$, since the equation $D_i \tilde g_{i0} = 0$ yields: 
$$\partial_0 (w^2(t,r)v(t,r)/u(t,r))=\partial_0 (L^4(t,r))=0,$$
i.e. $L(t,r)=L(0,r)\equiv L(r)$ has no time dependence. As for~$K(t,r)$, using the RTG equations~\cite{kni1, kni2}, namely the equation for~$R_{11}$, we obtain (for short, we omit the arguments $t$ and $r$ of all functions and use the dot notation for the time derivative and the prime notation for the derivative on the radius $r$):
\begin{equation}
\ddot K=\Bigg[-K^7 \Big(L L' r \big(2 e {s}^2+\mu^2/2\big)-4
K'^2 L^2 \mu^2+2 (L')^2 \big(\mu^2/2 r^2-1\big)\Big)+
\label{dynk}
\end{equation}
$$
+2
K^6 K' \Big(L^2 r \big(2 e {\eta}^2+\mu^2/2\big)+
L
\big(L' \big(4 \mu^2/2 r^2-5\big)+L'' r\big)-5
L'^2 r\Big)+
$$
$$+2 K^9 (L')^2 \mu^2/2-8 K^8 K' L L'
\mu^2/2+
$$
$$+2 K^5 K' L \Big(15 K' L' r
-2 L \big(K'
\big(2 \mu^2 r^2/2-3\big)+K'' r\big)\Big)-
$$
$$
-
24 K^4 K'^3
L^2 r+K^3 L^5 L' \mu^2 r/2-2 K^2 K' L^6 \mu^2 r/2+
$$
$$
+2 K
{\dot K} L^5 r (5 {\dot K} L'-2 {\dot K}' L)-16 {\dot K}^2 K'
L^6 r\Bigg]/\Big(4 K L^5 r (K L'-2 K' L)\Big)
$$
(here the graviton mass~$\mu$ should be set equal to 0, the density of matter is denoted by $e(t,r)$, the first component of its velocity $dx^1/ds$ --- by $\eta(t,r)$). The dynamics of matter can be obtained from the conservation law for the energy-momentum tensor~$\nabla_i T^{ij}=0$. Two nontrivial equations (for $j=0$ and $j=1$) take the form
$$
A_1\partial_0 e(t,r)+A_2\partial_0 \eta(t,r)+A_3 =0,
$$
(expressions for $A_i$ are too 
cumbersome to give them here in explicit form). From these equations one can obtain $\partial_0 e(t,r)$ and $\partial_0 \eta(t,r)$. The initial conditions for $e(t,r)$ corresponds to the static solution $e(0,r)=e(r)$, $\eta(0,r)=0$.

\begin{figure}
\includegraphics{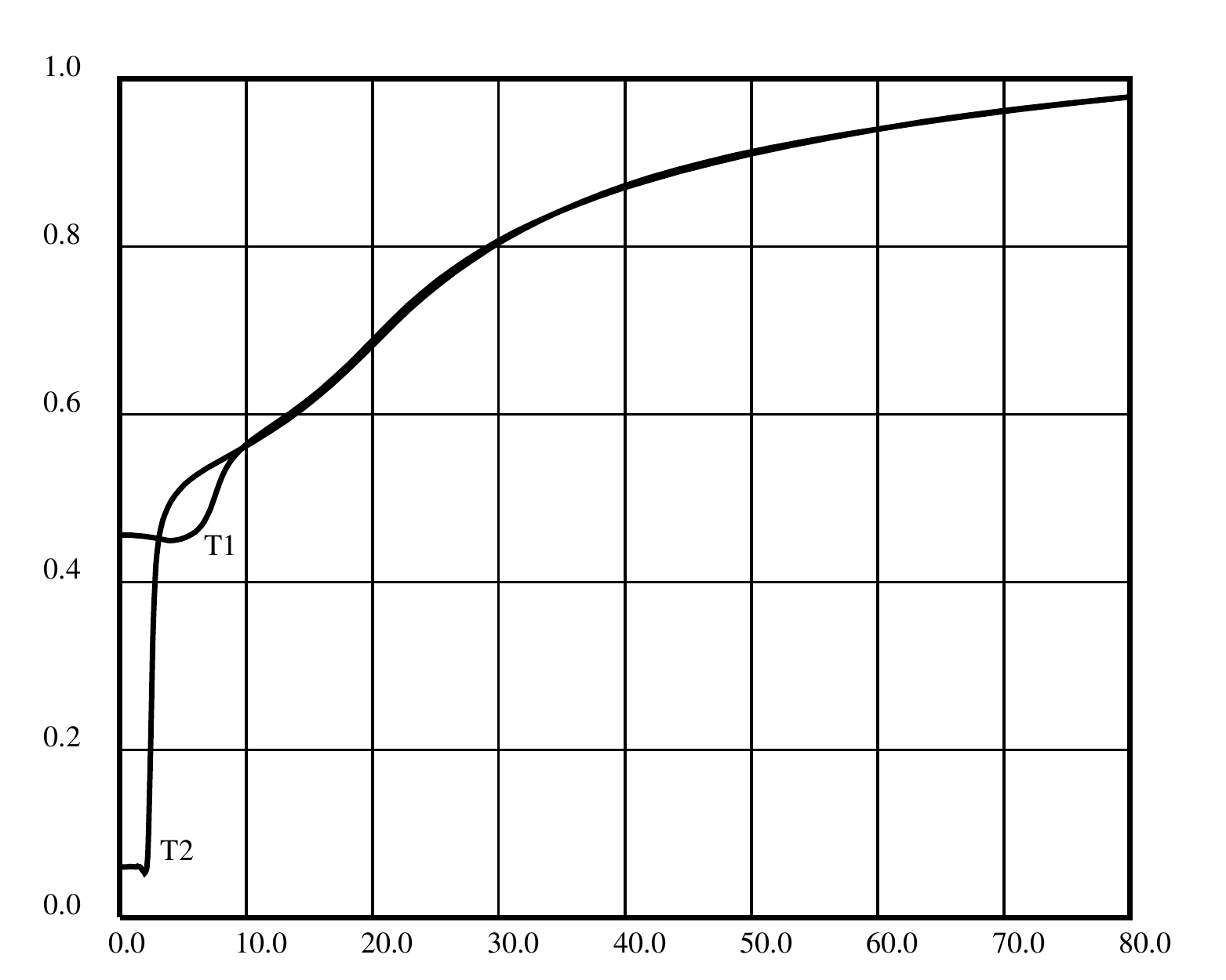}
\caption{Evolution of the metric coefficient~$u(t,r)$ at sufficiently large times ($T2>T1$). 
}
\end{figure}

The results of the numerical calculations confirm our presumptions: the collapse of cold dust~($C_1=0$) proceeds somewhat faster than the collapse of the matter with residual pressure proportional to~$C_1$~(of course, $C_1$ should be small enough, otherwise the collapse will not be possible for the 
the given mass of the solution). 
On Fig.~2 one can see the typical evolution of the metric coefficient~$g_{00}=u(t,r)$ at sufficiently large times.

As we already mentioned, there is no "boundary of body" (the matter is distributed continuously), but the boundary of the region in which most of the matter is concentrated coincides approximately with the region of rapid growth of $u(t,r)$ (see Fig. 2.).
Let us note that according to Birkhoff's theorem~(and its analogue in the RTG~\cite{birk, birk2}) the metric coefficients don't evolve in the region where the density of matter is sufficiently small~(i. e., in approximate vacuum).
It should be mentioned, that the velocity of the "boundary" decreases in the course of time, while the decrease of~$u(t,0)$ doesn't become slower. These two facts completely describe the subsequent evolution of the solution --- the boundary stops, while $u(t,0)$ continues to decrease.
This picture of collapse is in full agreement with the well-known Oppenheimer-Snyder solution \cite{oppen1}, \cite{oppen2} described from the point of view of the distant observer.

\section{Dynamics in Massive RTG }

Let us recall that for static solutions in massive RTG we should confine ourselves with the graviton mass values that are only a few orders of magnitude less than the mass of the solution.
For the dynamics in massive RTG we meet the same problem: as it will be shown, the effect of the non-zero graviton mass on the dynamics of the collapse becomes significant only after a time proportional to~$1/\mu$, i. e.~$10^{30}$, and it is impossible to perform calculations on such a time scale. So we should confine ourselves only with the graviton mass values that are a 3-5 orders of magnitude less than the mass of the solution.

The initial conditions correspond to the static solution of massive RTG (see Sec. 1.). Start of evolution (\ref{zapu}) and dynamical equations (\ref{dynk}) are precisely the same as in previous section, but the graviton mass $\mu$ now isn't equal to zero.
\begin{figure}
\includegraphics{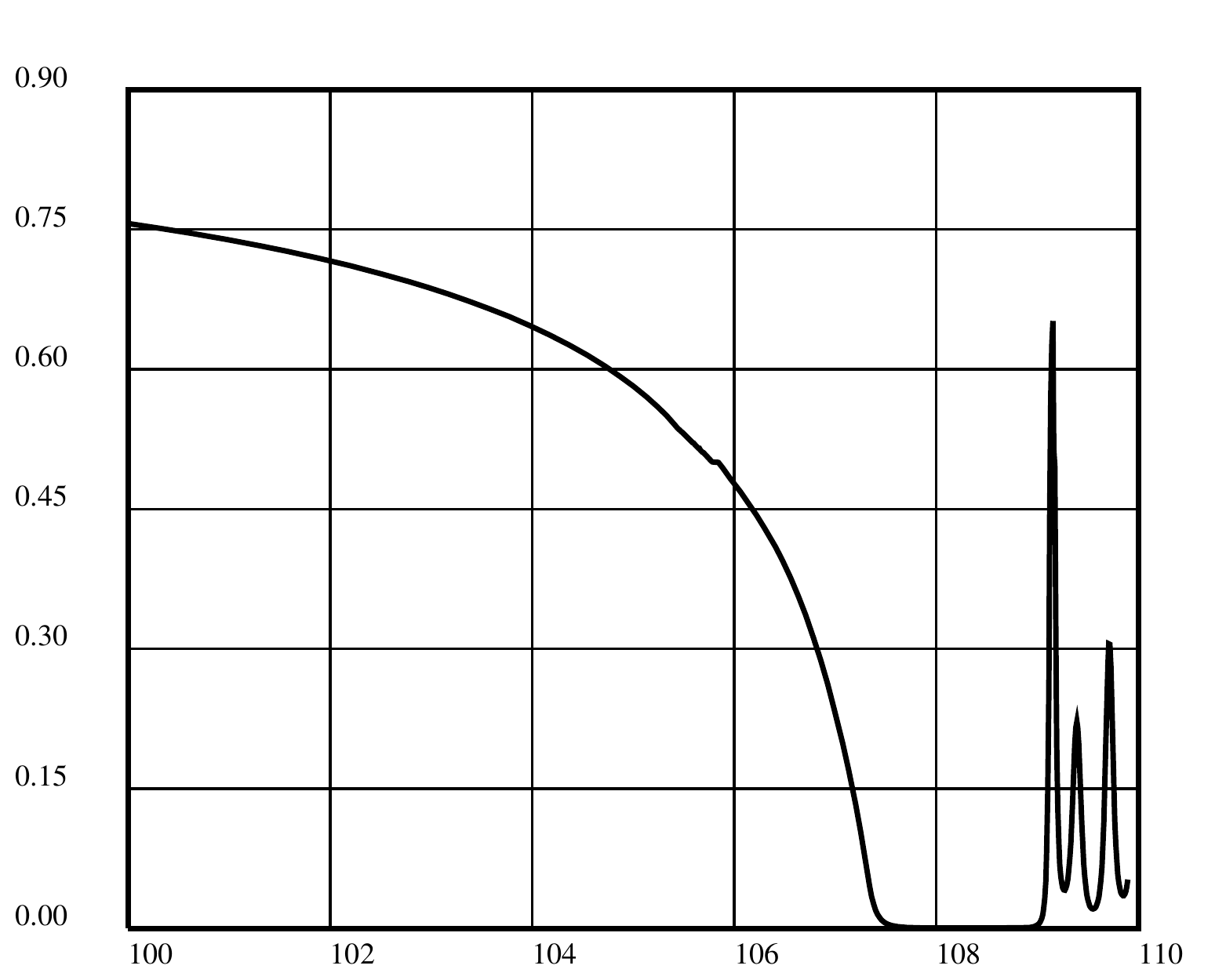}
\caption{Time dependence of the metric coefficient $u(t,0)$ before and after the turning point $t_0=108.81$.
}
\end{figure}

The comparison of the results of numerical simulations for the massive and massless cases leads us to the conclusion that at early stages the non-zero
graviton mass have no influence on the collapse dynamics. Only at sufficiently large times, when the metric coefficient $u(t,0)$ appears to be sufficiently small, the graviton mass changes the system dynamics drastically. The value $u(t,0)$ reaches a turning point and starts to increase. The smaller is $\mu$, the greater is the time of the turning point, when contraction is replaced by the expansion.
The subsequent evolution shows oscillations of~$u(t,0)$ (see Fig. 3) accompanied by irregular oscillations (waves) in the distribution of matter (see Fig. 4). It should be noted that the oscillations of $u(t,0)$ are not harmonic, and, similarly, the oscillations of density are not just uniform cycles of contraction/expansion~(see Fig.~3,4).
The initial value of a characteristic radius of static solution is never recovered in the course of these oscillations, i.e. the values of density at large $r$ during the process are considerably smaller than the initial ones.

\begin{figure}
\includegraphics{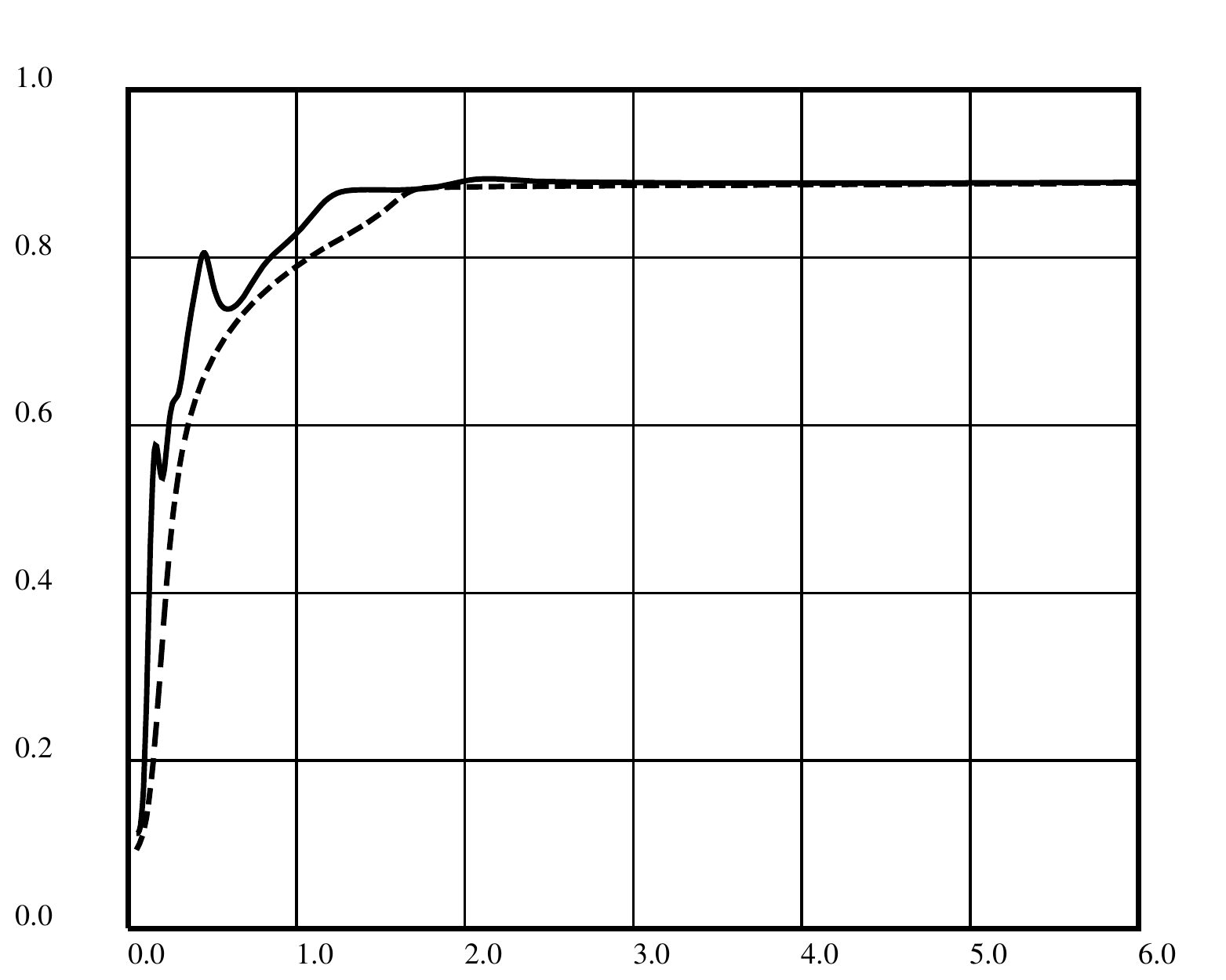}
\caption{Metric coefficients $u(t_1,r)$ and $u(t_2,r)$ as functions of $r$: at $t_1<t_0$ --- dashed line; at $t_2>t_0$ --- solid line; $t_0$ --- the turning point time. 
}
\end{figure}

The obtained results serve only as an indication of the main dynamics properties of the system considered. We should now to give a direct analytical proof that for~$\mu\not=0$ there is a turning point in the dynamics of $u(t,0)$, while at $\mu=0$ there is not. At the same time we will be able to obtain the dependence of the turning point time on the mass of the graviton.
Since we are considering a small neighborhood of~$r=0$, we use expansions on powers of~$r$.
Taking into account the boundary conditions at~$r=0$, for the functions $K(t,r)$ and $L(r)$ one can obtain
$$
K(t,r)=K_1(t)r+K_2(t)r^2+K_3(t)r^3+\ldots \qquad L(r)=L_1r+L_2r^2+L_3r^3 +\ldots
$$ 
Substituting these expansions into~(\ref{static}) and (\ref{dynk}) yields $K_2(t)=0$ and 
$L_2=0$.
Similarly,
$$
e(t,r)=e_0(t)r^0+e_1(t)r^1+e_2(t)r^2+\ldots \qquad \eta(r)=\eta_1(t)r+\eta_2(t)r^2+\eta_3(t)r^3 +\ldots
$$ 
Substituting these expressions into~(\ref{dynk}) and collecting terms with~$r^0$, we obtain (for short, we omit the argument~$t$ of  functions~$K_1(t)$ and $K_3(t)$):
\begin{equation}
\ddot K_1=-\frac{K_1^7 \mu^2}{4 L_1^4}-\frac{5 K_1^5
{L_3}}{L_1^5}+\frac{K_1^5 \mu^2}{8 L_1^4}+\frac{10
K_1^4 K_3}{L_1^4}+\frac{5 \dot K_1^2}{2
K_1}+\frac{K_1 \mu^2}{8} \label{full}
\end{equation}
During the collapse the coefficient $K_1(t)$ tends to zero, so in (\ref{full}) we can take into account only the terms linear in~$K_1(t)$ and neglect the terms with $K_1^4(t)$, $K_1^5(t)$, $K_1^7(t)$.
So we obtain:
\begin{equation}
\ddot K_1=\frac{5 \dot K_1^2}{2
K_1}+\frac{K_1 \mu^2}{8}
\label{line}
\end{equation}

It should be noted, that the solution of this equation (this equation is valid only for small $r$
and only for the late stages of collapse),
coincides with the solution of the equation arising in the cosmological model RTG \cite{cosm2}, obtained for the case of zero pressure and with assumptions, that the maximum value of the scale factor of the universe $ R_ {max} $ is sufficiently large. So 
at a small vicinity of the origin and for the late stages of collapse we (in some sense) can observe a ``homogeneous isotropic universe''. To evaluate the time of the turning point we note that the equation for the scale factor of the universe $ R (\tau) $ at the vicinity of the lower turning point 
$ R_ {min} $, obtained in \cite{cosm1}, coincides  with (\ref{line}).

For $\mu\not=0$ the solution of (\ref{line}) is:
\begin{equation}
K_1(t)=\frac{const}{\cos^{2/3}\big(\sqrt{3}\mu(t-t0)/4\big)},
\label{linemass}
\end{equation}

For $\mu=0$:
\begin{equation}
K_1(t)=\frac{const}{(t-t0)^{2/3}}
\label{linezer}
\end{equation}
From~(\ref{line}) it follows that for $t\ll 1/\mu$ the second term in the right hand side of equation~(\ref{line}) can be neglected, and the evolution of the system is described by expression~(\ref{linezer}).
For $t\sim 1/\mu$ the mass term becomes significant, the turning point is reached, $K_1(t)$ starts to increase and its behavior in the neighborhood of the turning point is described by equation~(\ref{linemass}).
Numerical simulation shows that after the turning point the quantity~$K_1(t)$ continues to increase, and after some time the linear approximation~(\ref{line}) becomes invalid~(see Fig. 3). 
In this process $K_1(t)$ reaches values comparable to its initial value, $K_1(0)$, and then starts to decrease again.
Therefore, we have proved that, in contrast to the massless theory, for arbitrarily small mass of the graviton the collapse stops, and a characteristic time elapsed from the beginning of the collapse to the turning point is proportional to~$1/\mu$.

\section{Conclusions}

The results of our numerical analysis performed for a wide class of model equations of state for the collapsing matter, on the one hand, confirm an intuitively obvious assumption that the presence of non-zero pressure slows down the process of the collapse, but doesn't result in any qualitative changes in the dynamics of the system. However, we should note, that we have considered relatively small values of pressure at which neither collapse stopping nor critical phenomena~\cite{krit} occur.
On the other hand, for the case of the cold dust model the status of the rigorous analytical solution of the general relativity equations has been confirmed in the framework of the massless RTG. 
The dynamics described by equations of the massless RTG, as it was expected, exactly coincides with the dynamics of general relativity solution taken in the reference frame of a distant observer. Namely, the coefficient~$g_{00}$ exponentially decays in the region where the density of matter is relatively large~(it is this 
circumstance that leads to unbounded growth of~$T^{00}$~\cite{kni2, rost} for finite matter density~$e(t,r)$), but reaches zero value only after infinite time.

We also have studied numerically the influence of non-zero graviton mass on the properties of static solutions for a wide class of model equations of state. It appears, that the increasing of graviton mass leads to
increasing of metric coefficient $g_{00}$ at the center of solution. So for a given equation of state and for the given mass of solution
the increasing of graviton mass leads to a ``more flat'' static solution. 

And, finally, we have studied the dynamics of the collapse in the massive RTG analytically and numerically. It appears that the time of a turning point in the collapse is proportional to $1/\mu$, where $\mu$ is the graviton mass. If we take into account the existing estimates for the graviton mass~$m_g<3\cdot 10^{-66}\;$g, we can conclude that this time is proportional to $4\cdot 10^{17}\hbox{sec}=12\cdot 10^{9}\;\hbox{years}$, i.e. it is comparable with the age of the Universe.
Due to technical constraints we have had to study a model in which the mass of the solution differs from the graviton mass by only 3--5 orders of magnitude. Nevertheless, the obtained results confirm conclusions of the previous analysis made in Ref.~\cite{dyn}. It should only be added that the pulsations of the dust sphere are irregular and its original size is not recovered in the course of evolution.

We are grateful to M. A. Mestvirishvili and K. A. Sveshnikov for their
attention to this work and fruitful discussions.


\begin{thebibliography}{99}
\bibitem {kni1} {\it A.A. Logunov, M.A. Mestvirishvili}. Relativistic Theory of Gravity [in Russian]. Moscow, Nauka, 1989.
\bibitem {kni2} {\it A.A. Logunov}. Relativistic Theory of Gravity [in Russian]. Moscow, Nauka, 2012.
\bibitem {stat} {\it A.A. Logunov and M.A. Mestvirishvili}. // Theor. Math. Phys. 1999. V. 121. P. 1262.
\bibitem {dyn} {\it S.S. Gershtein, A.A. Logunov, M.A. Mestvirishvili}. // Theor. Math. Phys. 2009. V. 161. P. 1573.
\bibitem {oppen1} {\it R.C. Tolmen}. // Proc. Nat. Acad. Sci. USA, 1934. V. 20. P. 169.
\bibitem {oppen2} {\it J.R. Oppenheimer, H. Snyder}. // Phys. Rev. 1939. V. 56. P. 455.
\bibitem {fock} {\it V.A. Fock}. Theory of space, time and gravitation. Moscow, GIITL, 1955.
\bibitem {avak} {\it R.M. Avakyan}. Precise solutions to Einstein equations and their 
interpretation. Tartu, Tartu University, 1988. P. 22.
\bibitem {genk} {\it A.V. Genk}. // Theor.Math.Phys. 1991. V. 87. P. 130.
\bibitem {birk} {\it A.A. Logunov, M.A. Mestvirishvili}. // Phys. Part. Nucl. 2009. V. 40. P. 67.
\bibitem {birk2} {\it A.A. Logunov, M.A. Mestvirishvili}. // Theor. Math. Phys. 2014. V. 181. P. 1471.
\bibitem{cosm2} {\it S.S. Gerstein, A.A. Logunov, M.A. Mestvirishvili}. //
Phys. Atom. Nucl. 1998. V. 61. P.1420;
Yad. Fiz. 1998. V. 61. P. 1526.
\bibitem {cosm1} {\it M.A. Mestvirishvili, Yu.V. Chugreev}. Theor. Math. Phys. 1989. V. 80, No. 2. P. 305.
\bibitem {rost} {\it A.A. Logunov, M.A. Mestvirishvili}. Theor. Math. Phys. 2013. V. 174. P. 253.
\bibitem {krit} {\it C. Gundlach, J.M. Martin-Garcia}. Living Rev. Rel. 2007. V. 10, No. 5.

\end{thebibliography}
\end{document}